\documentclass[letterpaper]{article}
\usepackage{aaai25}
\nocopyright
\usepackage{times}
\usepackage{helvet}
\usepackage{courier}
\usepackage[hyphens]{url}
\usepackage{graphicx}
\usepackage{natbib}
\usepackage{caption}
\usepackage{amsmath}
\usepackage{amssymb}
\usepackage{booktabs}
\usepackage{xcolor}

\title{Two Fault Lines: Latent Polarity Geometry in X Community Notes}
\author{Andreas Andreou, Michael Sirivianos}
\affiliations{Cyprus University of Technology, Limassol, Cyprus\\
andreas.i.andreou@cut.ac.cy, michael.sirivianos@cut.ac.cy}

\begin{document}
\maketitle

\begin{abstract}
Community Notes is X's crowdsourced fact-checking system. A note is published beneath the post it corrects, only when raters who usually disagree both rate it helpful, a design called bridging. To apply that rule, the system learns who disagrees with whom from the ratings alone, placing every rater and note on one line, the polarity axis, so that raters who judge notes alike sit close together. Every scorer in the production pipeline represents a viewpoint on a single axis. Refitting the base model these scorers share on the full public data (212.9M ratings, 2.33M notes, 1.07M raters), we find that one axis is too few. The space is \textbf{at least two-dimensional}. The first axis is left/right politics, while the second, which we interpret as trust in institutions, is largely independent of the first, and a held-out test confirms that the second axis improves prediction of ratings the model has not seen, while a third axis adds little.  A second rater dimension learned from one set of topics predicts how raters judge COVID and Ukraine notes excluded from the fit, so it does not merely restate subject matter. Second-axis disagreement predicts non-publication. Among heavily rated notes that barely divide raters politically, the published share falls from 71.5\% to 11.7\% as second-axis disagreement grows. A one-axis fit records these notes only as weakly polarised and less helpful; the information that raters at one end of the second axis support them is lost. Authors write notes matching their own position on both axes (r = 0.538 and 0.358), and a small minority of raters cast most ratings (Gini = 0.718). Fewer notes are published in the smallest language communities, but the shortfall is in ratings received, not in how the rule treats them. Keeping ratings per note constant, only Hindi stays below the global rate of 10.85\%, and Greek moves from 7.76\% to 11.68\%. We argue for a bridging model that allows more than one axis of disagreement, and for recruiting raters in the languages the current design reaches least.
\end{abstract}


\section{Introduction}

Social media platforms have struggled to moderate mis/disinformation without appearing to take a side \citep{gillespie2018}. False news spreads more widely online than true news \citep{vosoughi2018}, exposure to it is concentrated in a small minority of accounts \citep{grinberg2019}, and what to do about it has become a research field in its own right \citep{lazer2018,allcott2017}. X's answer is Community Notes, a crowdsourced fact-checking system in which ordinary users, not platform staff, write short notes that correct or add context to misleading posts \citep{wojcik2022,prollochs2022}. It differs from an ordinary up-vote system in one rule. A note is shown publicly only once raters who normally disagree with each other have both judged it helpful. Published notes reduce resharing of the posts they annotate \citep{chuai2026}. The same ``bridging'' logic is now being proposed for recommender systems and governance problems well beyond fact-checking \citep{ovadya2023}.

To apply the rule, the system must infer who disagrees with whom, which it does from the ratings alone. It fits a regularised latent-factor model, the matrix factorisation standard in recommender systems \citep{koren2009}, adapted for bridging \citep{wojcik2022}. Raters who judge the same notes the same way receive nearby factor values, and raters who judge them differently receive distant ones. The fitted model gives each rater and each note one scalar, and the relative position of a note and its raters on that scalar decides publication.

This paper tests, at the scale the system runs, whether disagreement about a post fits on one line. Two raters can share a political position yet differ in how far they trust institutions such as public health agencies, mainstream news outlets and official statistics. A note both endorse has bridged a real disagreement. If the two raters occupy the same point on the single scalar, the model records no bridge and the note may not be published.

The second failure mode concerns rater numbers, not geometry. The rule cannot fire at all unless a note attracts ratings from raters on both sides, and in a small language community, the pool may never supply them. A note can go unpublished because the model cannot represent the disagreement it bridges, or because too few raters arrived to demonstrate any bridge. Neither failure is visible in the platform's own output, and we measure both.

Prior work has largely taken the one-dimensional design as given, validating the published factor against external ideology measures or measuring downstream effects (Section~2). Whether the disagreement the system is designed to bridge is itself one-dimensional has not been tested. We test this by asking whether the platform's bridging algorithm discards a second axis of disagreement.

In this paper, we answer three research questions concerning the bridging model.

\begin{enumerate}
\item \textbf{RQ1 (Dimensionality).} Is the disagreement Community Notes is designed to bridge one-dimensional? If there is more than one independent axis, a one-dimensional model throws information away and will systematically fail to publish notes that bridge one divide while splitting on another.
\item \textbf{RQ2 (Author bias).} Do note authors write notes aligned with their own viewpoint? If so, how strong is that bias and what share of otherwise partisan raters nevertheless produce centrist, cross-camp notes?
\item \textbf{RQ3 (Language equity).} Does the one-dimensional bridging mechanism work equally well across language communities? Language technology and content moderation already underserve low-resource languages \citep{joshi2020,blasi2022}. Small-language communities face an additional constraint, since a rater pool below a certain size cannot supply the cross-camp ratings that publication requires.
\end{enumerate}

We answer all three by refitting the base bridging model the platform's scorers share on the full public rating data (212.9M ratings; snapshot 2026-06-14), extending it from one dimension to two. Our contributions are:

\begin{itemize}
\item \textbf{The first full-scale reconstruction, to our knowledge, of the Community Notes polarity geometry.} We show that the space is at least two-dimensional, with a second axis, distinct from left/right politics, that we interpret as institutional trust. We establish this through out-of-sample prediction rather than variance share, since a large share alone is not evidence of a real extra axis. We also show that a second rater dimension learned from one set of topics predicts ratings on topics excluded from that fit. This rules out two alternative readings, that the dimension restates a note's subject matter and that it restates how contested the note is. It does not establish that the transferring dimension is the institutional-trust axis.
\item \textbf{A measurement of what a one-axis fit leaves out.} Among notes that barely split the first axis and carry at least 100 ratings, X publishes 71.5\% of those nearest the second-axis origin and 11.7\% of those furthest from it. Both fits give these notes the same low note intercept, the term the publication rule uses, so the gap is not an artefact of fitting one axis, and some of these notes may be less helpful on their merits. What the rank-1 fit loses is the second coordinate, which records that raters at one end of the second axis rate them helpful; it reduces that support to a below-average factor and sees only a weakly polarised note of lower appeal. Whether such notes should publish is a design choice we leave open (Section~5.1).
\item \textbf{A full-data measurement of the mechanism's author bias, rater concentration, and language-coverage gaps}, with concrete implications for system design and for information-integrity monitoring more broadly.
\end{itemize}

\section{Background and Related Work}

\subsection{The Community Notes model}

Community Notes (formerly Birdwatch) is X's crowdsourced fact-checking system, launched publicly in 2021. Any user meeting activity thresholds can become a Community Notes contributor, writing notes that appear beneath posts and rating notes written by others. A note is publicly shown (``Currently Rated Helpful'') only when it achieves cross-partisan support, the central design principle intended to prevent capture by a single ideological group \citep{wojcik2022}. It is the largest deployment of crowdsourced fact-checking to date \citep{saeed2022}.

The algorithmic backbone is a bridging-based matrix factorisation model, introduced for Birdwatch by \citet{wojcik2022} and described in X's open-source implementation. The same idea was later generalised as bridging-based ranking \citep{ovadya2023}. The rating prediction model is:

\begin{equation}
\hat{r}(u,n) = \mu + i_u + i_n + \mathbf{f}_u \cdot \mathbf{f}_n
\end{equation}

where $\mu$ is a global intercept, $i_u$ and $i_n$ are per-rater and per-note intercepts. The note intercept $i_n$ indicates how helpful a note is considered irrespective of viewpoint, and it is the term the publication rule uses. The vectors $\mathbf{f}_u, \mathbf{f}_n$ are latent factors capturing viewpoint alignment. The factor is easiest to read geometrically, where each rater and each note is a point on an axis, with zero (the \textit{origin}) as the neutral midpoint. The factor's sign says which side of the origin the rater or note falls on; we call the two sides \textit{camps}. The dot product $\mathbf{f}_u \cdot \mathbf{f}_n$ is positive when the rater and note share a side, and negative when they do not. The publication threshold ensures that a note must receive sufficiently high predicted ratings from raters in \textit{both} camps. This bridging mechanism is the primary innovation over earlier crowdsourced fact-checking designs \citep{kim2019}.

The functional form is the standard latent-factor model of recommender systems, where user and item factors are learned jointly from sparsely observed ratings \citep{koren2009}. What differs is how the factors are used. A recommender uses the dot product as a similarity score, serving each user the items closest to their own position. Bridging rewards agreement \textit{across} positions \citep{ovadya2023} instead. The dimensionality of the factor space is a tuning parameter for a recommender, but here it fixes how many kinds of disagreement the publication rule can see, which makes it a design choice about the mechanism.

The official X implementation keeps the factor one-dimensional (scalar $f_u, f_n \in \mathbb{R}$) ``to avoid overfitting'' and expects to raise its dimensionality as the data grow \citep{xcn2026}. The deployed pipeline adds Expansion, Topic and per-group models and a Gaussian aggregation, each representing a viewpoint with a single factor, together with a factor gate ($|f_n| < 0.5$), net-helpful minimums, rater-helpfulness filtering and tag filters. We refit the base model these scorers share and extend it to 2-D ($\mathbf{f}_u, \mathbf{f}_n \in \mathbb{R}^2$) to test whether a single axis of disagreement adequately characterises the data.

\subsection{Latent ideology models in political science and social media}

Latent factor models to estimate political ideology have a long history. In legislative roll-call analysis, W-NOMINATE and related models \citep{poole1985,clinton2004} recover low-dimensional ideology spaces from voting records, where a single left-right dimension accounts for most variance, but a second adds explanatory power in some legislatures and periods. Two lessons are relevant to our setting.  First, the number of dimensions is treated there as something to be established from the data, and studies that ask the question find the answer varies by legislature and period \citep{hix2006}. Second, recovered low dimensionality is not necessarily a fact about underlying preferences: \citet{aldrich2014} show that when parties are strongly polarised, votes line up along the party split and a one-dimensional summary fits well even if legislators differ on several underlying issues. Low recovered dimensionality can therefore reflect the voting environment rather than the preferences themselves. A bridging system that fixes the dimension at one by design inherits this risk, and cannot detect it from its own output.

Applied to social media, analogous latent models have estimated ideology from following networks \citep{barbera2015}, retweet patterns \citep{conover2011,barbera2015b}, co-engagement with news sources \citep{bakshy2015}, and a joint factorisation of network structure and content consumption \citep{lahoti2018}; almost all report a single dominant partisan score. Community Notes is a distinctive setting for this inquiry, since the factors are estimated from explicit evaluation events, in which a rater reads a note and rates it helpful or not, rather than from network structure. That avoids the homophily bias of co-engagement measures, where network position proxies social circles as much as beliefs, so the factors are closer to disagreement about specific claims than network measures are.

\subsection{Crowdsourced fact-checking: effectiveness and limitations}

Crowdsourced fact-checking responds to two limits of the professional kind: effectiveness and coverage. Fact-checking moves political beliefs with a modest overall effect ($d = 0.29$) across 30 studies \citep{walter2020}. Coverage is the key limit here. Professional fact-checking cannot reach everything, so most false content stays unlabelled, and \citet{pennycook2020} show this is not neutral. Attaching warnings to some false headlines makes those that escape a warning look \textit{more} accurate, an implied truth effect. A system that covers some communities well and others barely at all can therefore make uncovered communities' unlabelled content look verified. The theoretical basis for the crowdsourced alternative is that laypeople judge headline accuracy reasonably well \citep{pennycook2021}, that a small, politically balanced crowd can match professional fact-checkers on contested political news \citep{allen2021}, and that crowd credibility ratings track journalism experts more closely than science experts \citep{bhuiyan2020}.

Community Notes specifically has been studied from several angles. Raters judge the same biased content differently depending on their own partisan lean  \citep{allen2022}. Notes are rated more helpful when they cite trustworthy sources \citep{prollochs2022}. In a later study, the odds of a helpful judgement are 2.33 times higher with an external source link, and notes linking to high-bias sources of either political side are judged significantly less helpful \citep{solovev2025}. Two recent studies raise a sharper concern about manipulability: \citet{truong2025}, in simulation, find that the algorithm suppresses a substantial share of helpful notes and is highly sensitive to rater bias, with small coordinated groups able to manipulate which notes appear; and \citet{alimohammadi2026} argue that consensus-based moderation pressures minority raters toward the majority view, proposing that raters be weighted by demonstrated reliability rather than by agreement.

Closest to the present work, \citet{bouchaud2026} cross-reference the platform's own one-dimensional latent factor against external ideological scales across 13 countries, finding that it aligns with a Left-Right axis and that the system systematically undermoderates polarising content around recent elections. Their analysis treats the factor as given and one-dimensional. Less attention has gone to the \textit{geometry} of the latent polarity space when the model is allowed more than one axis: whether the camps are separated or overlapping, whether one dimension suffices, and where different topics sit. These questions decide which notes get published and which communities go uncovered; we examine them by refitting the factorisation at higher rank rather than validating the platform's scalar against outside data.

\subsection{Multilingual coverage gaps}

Content moderation and automated content-analysis systems underperform in non-English languages \citep{nicholas2023}: training data and model development concentrate in a few high-resource languages \citep{joshi2020}, and \citet{blasi2022} find the resulting inequality systematic. Community Notes adds a structural requirement on top. A note must accumulate ratings from ideologically diverse raters, and in small-language communities the rater pool may be too thin even when ideologically representative. Closest to this question, \citet{stewart2025} document that Community Notes participation is highly uneven across languages and that reliance on professional fact-checking remains limited; we ask instead how the bridging-factor model performs when the rater pool is thin. We take Greek-script content as the case, a small-language public sphere with documented exposure to online disinformation \citep{giomelakis2024}.

\section{Data and Model}

\subsection{Data}

We use the public Community Notes data snapshot from 2026-06-14, retrieved from X's public data release at \url{https://x.com/i/communitynotes/download-data} (8 rating shards plus note text and \texttt{noteStatusHistory} publication decisions). Before fitting, we iteratively filter to notes with at least 5 ratings and raters with at least 10 ratings, repeating until stable, so that every retained factor is estimated from a minimum amount of data. After filtering, the fitted dataset comprises \textbf{2,334,630 notes} and \textbf{1,067,909 raters}; we fit the model on \textbf{212,900,053 ratings}, holding out a further 2\% for validation. All data are publicly released by X under the Community Notes data licence.

\subsection{Factor model}

We fit the bridging model with a 2-D factor extension:

\begin{equation}
\hat{r}(u,n) = \mu + i_u + i_n + \mathbf{f}_u \cdot \mathbf{f}_n, \quad \mathbf{f}_u, \mathbf{f}_n \in \mathbb{R}^2
\end{equation}

The model has four parameter blocks: a global intercept $\mu$; a per-rater intercept $i_u$ (one number per rater); a per-note intercept $i_n$ (one number per note); and the factor matrices $\mathbf{F}_u$, $\mathbf{F}_n$, which hold two numbers per rater and two per note, their positions on the two axes. Ratings are encoded numerically: ``Helpful'' $= 1$, ``Somewhat Helpful'' $= 0.5$, ``Not Helpful'' $= 0$.

\textbf{Regularisation.} We use L2 regularisation, which keeps the model from overfitting raters or notes with few ratings, with $\lambda_i = 0.15$ on intercepts and $\lambda_f = 0.03$ on factors, with squared errors summed rather than averaged over ratings, matching the X open-source implementation. Each rater's and note's penalty is therefore fixed while its data term grows with its ratings, so raters and notes with few ratings are pulled more strongly toward zero.

\textbf{Optimisation.} We use Adam \citep{kingma2015} with learning rate $0.05$ and default settings ($\beta_1 = 0.9$, $\beta_2 = 0.999$, $\varepsilon = 10^{-8}$) for 300 epochs over the full rating matrix. Ratings are processed as a sparse list of (note, rater, value) triples (via \texttt{torch.scatter\_add\_}), so the full rater-by-note matrix is never built in memory. The full 212.9M-rating fit runs on an NVIDIA T4 GPU (Google Colab); wall-clock time is approximately 3 hours for 300 epochs.

\textbf{Initialisation.} The per-rater and per-note intercepts are initialised to zero, and the global intercept $\mu$ to 0.5 (the midpoint of the 0/0.5/1 rating scale); the 2-D factor matrices are initialised with small Gaussian noise ($\sigma = 0.1$) to break symmetry between the two dimensions.

\textbf{Identifiability.} The 2-D factor solution is not rotationally identified, because any orthogonal rotation of $(\mathbf{F}_u, \mathbf{F}_n)$ yields the same predicted ratings, so the model has many equally good solutions that differ only in how the two axes are rotated. After fitting, we fix this rotation using the SVD: we compute the SVD of $\mathbf{F}_n = \mathbf{U} \mathbf{\Sigma} \mathbf{V}^T$ and set $\mathbf{F}_n \leftarrow \mathbf{U}\mathbf{\Sigma}$, $\mathbf{F}_u \leftarrow \mathbf{F}_u \mathbf{V}$. This ensures dim 1 is always the largest source of disagreement in the note factor matrix, which makes the two dimensions interpretable and comparable across runs. The two axes are close in scale (singular values 717 and 619), but the direction is nonetheless sharply determined. Bootstrapping the note set 400 times moves the principal axis by 0.14 degrees (SD), and the dim-2 topic ordering of Table~\ref{tab:topic-bridge} is unchanged in every resample.

\textbf{Proxy validation.} We cannot observe X's internal model directly, so we check our refit against the publication decisions that the model made. Our re-estimated note intercepts separate notes X published (Currently Rated Helpful) from notes it rejected (Not Helpful), with an area under the ROC curve (AUC) of $0.993$, where 1.0 is perfect separation, and 0.5 is chance ($0.880$ against all other statuses combined; Appendix A). The separation between published and rejected notes is driven by the intercept. The first factor alone does not separate the two classes (AUC 0.44). This is expected, because the publication rule uses the factor only to check agreement across camps, not to judge helpfulness. This check was run on an earlier one-dimensional fit over a smaller filtered set (1.57M notes, 672K raters), so it bears on the estimation procedure and the intercept, not the factor geometry of the two-dimensional fit reported below.

\subsection{Analysis methods}

\textbf{Variance decomposition.} We report each dimension's share of total note-factor variance.

\textbf{Bimodality.} A high bimodality coefficient means a distribution has two separate peaks, two camps rather than one continuous spread. We use Sarle's bimodality coefficient
\begin{equation}
BC = \frac{\text{skewness}^2 + 1}{\text{excess kurtosis} + \dfrac{3(n-1)^2}{(n-2)(n-3)}}
\end{equation}
where $BC > 0.555$ indicates bimodality \citep{pfister2013}. We compute BC separately for note and rater factors on each dimension.

\textbf{Topic assignment.} We assign notes to the 10 topic categories in Appendix B using keyword matching on the note summary text. Matching is case-insensitive; the first matching topic wins; unmatched notes are labelled ``Other.'' Each topic's mean position on the two axes drives the axis interpretation in Section~4.2.

\textbf{Author--note correlation.} For notes whose author is also a rater (i.e., appears in the rater factor table), we compute Pearson $r$ between the author's rater factor and the note's factor, on both dimensions. We also compute the same-camp rate (author and note on the same side of the dim-1 origin) and identify bridge authors (polarised raters with centrist note production).

\textbf{Greek-script detection.} We detect Greek-script notes using a Unicode range filter covering the Greek (U+0370--U+03FF) and Greek Extended (U+1F00--U+1FFF) blocks, applied to the note summary field.

\section{Results}

The results follow the three research questions: whether the polarity space is more than one-dimensional, what geometry it has and what that costs (RQ1, Sections~4.1, 4.2 and~4.5), whether note authors write to their own side (RQ2, Section~4.3), and how the mechanism serves small language communities (RQ3, Section~4.4). The proxy validation in Section~3.2 (details in Appendix~A) showed that the refit closely tracks the official publication decisions;  robustness, participation, and coverage analyses are in Appendices~C and~D.

\subsection{The polarity space is  at least two-dimensional}

\textbf{A second latent axis improves held-out rating prediction, so it reflects a real structure rather than an artefact of the fit.} We fit the model at ranks $K = 0$, 1, 2, and 3, where the rank $K$ is the number of factor dimensions each rater and note gets ($K = 0$ means intercepts only), on one fixed 98/2 train/validation split, holding out individual (note, rater) ratings rather than whole notes or raters, so every note and rater still appears in training. Error is root mean squared error (RMSE; lower is better) on the held-out 2\%, and training runs to a 400-epoch budget with early stopping on that split; $K = 0$, 1 and 3 stopped early, while $K = 2$ reached the budget, so its RMSE is an upper bound and its gain over $K = 1$ a lower bound (Table~\ref{tab:rmse-by-rank}).

\begin{table}[t]
\centering
\small
\begin{tabular}{lrr}
\toprule
Rank $K$ & Validation RMSE & Gain vs $K{-}1$ \\
\midrule
0 (intercepts) & 0.3771 & n/a \\
1 & 0.3085 & $+0.0687$ \\
2 & \textbf{0.2899} & \textbf{+0.0185} \\
3 & 0.2892 & $+0.0008$ \\
\bottomrule
\end{tabular}
\caption{Held-out validation RMSE by model rank, on one fixed 98/2 split. Gains are computed from unrounded values and may differ in the last digit from the differences of the rounded RMSEs shown. The second factor's gain is positive in every split; the third factor's is marginal.}
\label{tab:rmse-by-rank}
\end{table}

Adding the second factor lowers held-out RMSE by $0.0185$ on the 0/0.5/1 rating scale, a 6.0\% relative reduction and about a quarter of the gain the first factor buys; over five random 98/2 splits, the gain is $+0.022 \pm 0.008$ (range $+0.015$ to $+0.030$), never approaching the permutation-null floor below. The third factor adds only $0.0008$ ($0.3\%$ relative), roughly a twentieth of the second's contribution. The held-out comparison runs on the most recent public snapshot (2026-07-05; 55.7M ratings, 718K notes, and 531K raters after the same filtering), not the exact ratings behind the descriptive fit: X's public release retains only the newest snapshot and has since shrunk, so the 2026-06-14 rating file is no longer retrievable (Section~5.4). We had run the comparison up to $K = 2$ on those ratings before they became unavailable ($K = 0$ 0.3759, $K = 1$ 0.2936, $K = 2$ 0.2784; gain $+0.0152$), which confirms the second axis on the exact snapshot behind the descriptive fit; $K = 3$ was not completed there, which is why Table~\ref{tab:rmse-by-rank} is reported on 2026-07-05. The descriptive variance shares and topic factors below are from the primary 2026-06-14 fit.

\textbf{Permutation null.} As a direct check, we refit the $K = 2$ model on deliberately scrambled data, where every rater still rates the same notes, but the rating values are randomly reshuffled among the pairs, destroying any link between who rated what and how, while the shape of the data stays identical. The null shows no held-out improvement over $K = 0$ (gain $-0.0002$), yet its two dimensions still split variance 50/50. A second dimension absorbs about half the variance even from pure noise, so the dim-2 variance share is not by itself evidence of a second axis. The held-out gain is such evidence, because  the null does not reproduce it.

\textbf{Variance decomposition and dimensionality.} On the full 2026-06-14 data, the fitted factor variances split 57.4\% / 42.6\% between dim 1 and dim 2 (ratio $1.35\times$); an earlier fit on a 15\% subsample of the ratings split 52.0\% / 48.0\% (ratio $1.08\times$), so the full data separate the two axes more clearly. On the 2026-07-05 snapshot, where the $K = 2$ split is $53.8\%$ / $46.2\%$, $K = 3$ splits $36.1\%$ / $33.0\%$ / $30.9\%$, a dim2/dim3 ratio of $1.07$ that would suggest three axes of near-equal weight. Judged by held-out prediction, the data support two robust axes and, at most, a weak third. We therefore retain $K = 2$ as the primary model. Whether the small $K = 3$ gain reflects a third dimension or residual overfitting is left open; a full-scale $K = 3$ fit under the same held-out protocol would settle it.

\textbf{A second rater dimension transfers across topics.} The held-out test shows a second axis predicts ratings, but not that it reflects a viewpoint rather than the note's topic or how contested the note is. We test this with a leave-two-topics-out design. We refit the $K = 2$ model on the assigned-topic ratings while excluding the two topics that load most heavily on the second axis, COVID/vaccines and Ukraine/Russia (45.9M training ratings from the other eight topics), so each rater's factors are learned with zero exposure to either target. We then freeze the rater factors and, for each held-out topic, fit only the note factors on a random half of that topic's ratings and predict the other half; with the rater side held fixed, this reduces to a standard ridge regression with an exact, closed-form solution. If a rater's second-axis position, estimated from unrelated topics, predicts how they rate the target, the axis is a transferable viewpoint. Neither alternative explanation survives this test. Topic membership cannot produce the gain, since the factor was learned without ever seeing the topic. Contentiousness cannot either, because the model's prediction depends on the product of the rater's and the note's positions, so  it has to say \textit{which} raters will like a given note, not just that the note divides people.

\begin{table*}[t]
\centering
\small
\begin{tabular}{lrrrrr}
\toprule
Held-out topic & Eval & K0 & K1 & K2 & 2nd-axis \\
& ratings & intercepts & +1st axis & +2nd axis & gain vs null \\
\midrule
COVID/vaccines & 1.87M & 0.3835 & 0.3348 & \textbf{0.2705} & \textbf{+0.064} vs $-0.007$ \\
Ukraine/Russia & 2.03M & 0.3831 & 0.3517 & \textbf{0.2822} & \textbf{+0.070} vs $-0.006$ \\
\bottomrule
\end{tabular}
\caption{Leave-two-topics-out cross-topic transfer test: held-out RMSE on target-topic ratings, with rater factors learned with zero exposure to the target topic.}
\label{tab:transfer}
\end{table*}

The full two-dimensional rater subspace, learned without the target, lowers held-out RMSE by $0.113$ (COVID) and $0.101$ (Ukraine) over intercepts alone (Table~\ref{tab:transfer}); the second axis contributes $0.064$ and $0.070$ of that beyond the first. A permutation null that shuffles the second-axis rater values across raters yields no gain ($-0.007$ and $-0.006$), so the real second axis outpredicts a permuted one by $0.071$ and $0.075$: fitting note factors against shuffled rater factors recovers nothing, and the gain reflects rater-note alignment, not the extra free parameter. There is no leakage, as the held-out ratings never entered either the frozen rater factors (other topics) or the fitted note factors (the fit half). For these two topics, the second axis contributes more than the first axis  ($0.064$ vs $0.049$ for COVID; $0.070$ vs $0.031$ for Ukraine): a rater's second-axis position, learned from unrelated topics, predicts how they rate science-credibility and geopolitical-information notes. This is our most direct evidence that the second dimension is a topic-independent viewpoint rather than an artefact of the factorisation. When we rotate the transfer fit's rater factors to principal axes (using the same procedure as in Section~3.2) and correlate each with the corresponding factor from the main full-data fit, the transfer fit's dominant axis tracks the main fit's dim~1 ($r = 0.63$; signs are arbitrary), as expected. Its second axis, however, correlates similarly with dim~1 and dim~2 ($r = -0.26$ and $r = 0.26$, respectively). The transfer establishes a second rater dimension that generalises across topics, not that this dimension is specifically the institutional-trust axis of Section~4.2. Two caveats apply. The transfer fit excludes the target topics and uses the noisier assigned-topic subset (45.9M ratings), which may explain why the correspondence correlations are only moderate. The recovered rater subspace is also near-isotropic (singular values 541 and 489), so the rotation that separates its two columns is weakly determined. 

\subsection{Geometry: two camps and a gradient}

\begin{figure*}[t]
\centering
\includegraphics[width=0.62\textwidth]{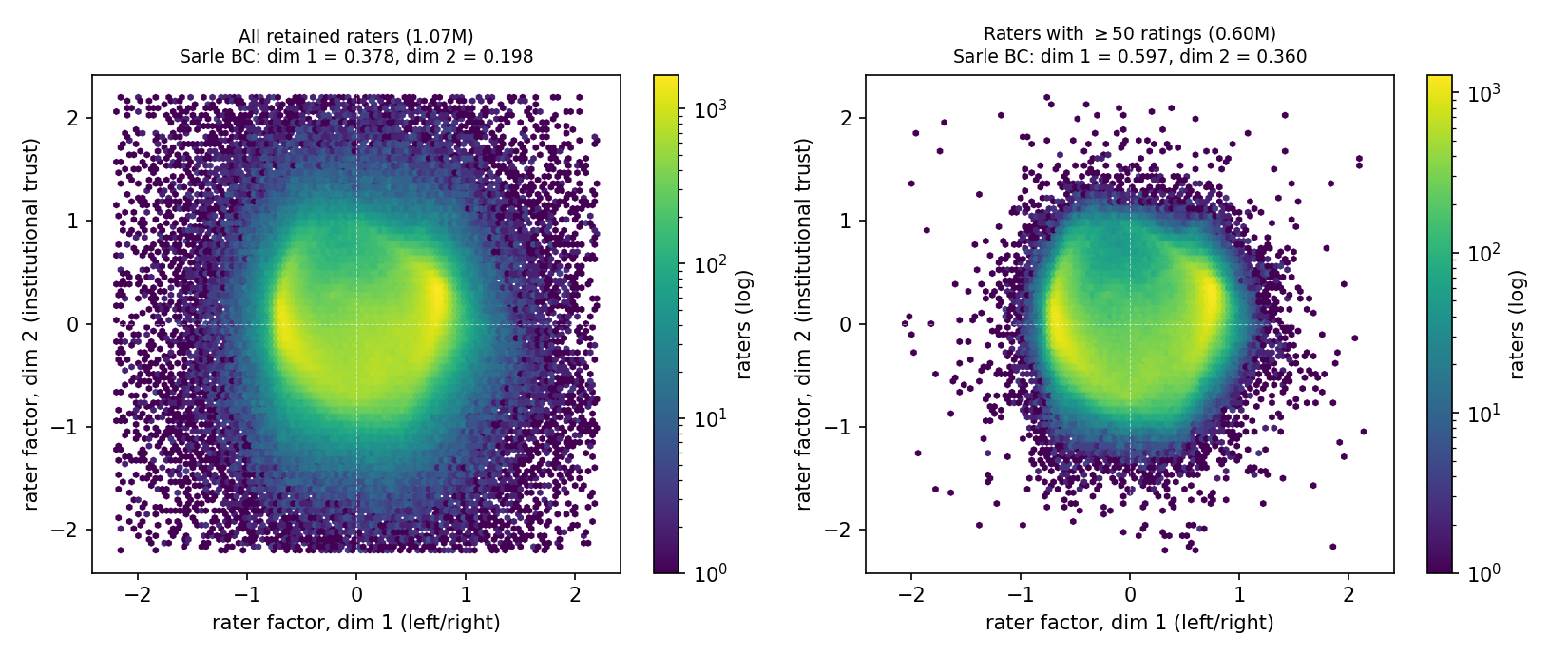}
\caption{Rater viewpoint density (hexbin, log colour scale). Left: all 1.07M retained raters; those with few ratings are pulled toward the origin by regularisation, producing one apparent peak and depressing Sarle's BC (0.378 dim 1, 0.198 dim 2). Right: raters with at least 50 ratings (0.60M); the origin mass is gone, two lobes appear along dim 1 ($BC = 0.597$, above the 0.555 bimodality threshold), and dim 2 stays continuous ($BC = 0.360$).}
\label{fig:space}
\end{figure*}

Bimodality depends on the activity of the rater. Across all retained raters, neither factor is bimodal by Sarle's criterion ($BC = 0.378$ on dim 1, $0.198$ on dim 2), because raters with few ratings carry little signal and regularisation pulls them toward the origin, inflating the central mass (Figure~\ref{fig:space}). Among raters the model can estimate, dim-1 BC rises monotonically with activity, crossing the $0.555$ bimodality threshold at 50 or more ratings ($0.597$, 95\% CI $[0.588, 0.604]$) and reaching $0.641$ past 100; dim 2 stays below the threshold at every activity level ($0.20$ to $0.38$). The note factors behave similarly but cross later. Dim-1 BC is $0.395$ across all retained notes, still straddles   the threshold at 50 or more ratings ($0.555$, 95\% CI $[0.553, 0.556]$) and clears it only past 100 ($0.597$, $[0.596, 0.599]$), while dim 2 stays continuous throughout (the full activity-filter table is in Appendix~C). Among raters with 50 or more ratings, the left/right axis is camp-like. The institutional-trust axis, in contrast, is a gradient at every activity level.

\begin{figure*}[tp]
\centering
\includegraphics[width=\textwidth]{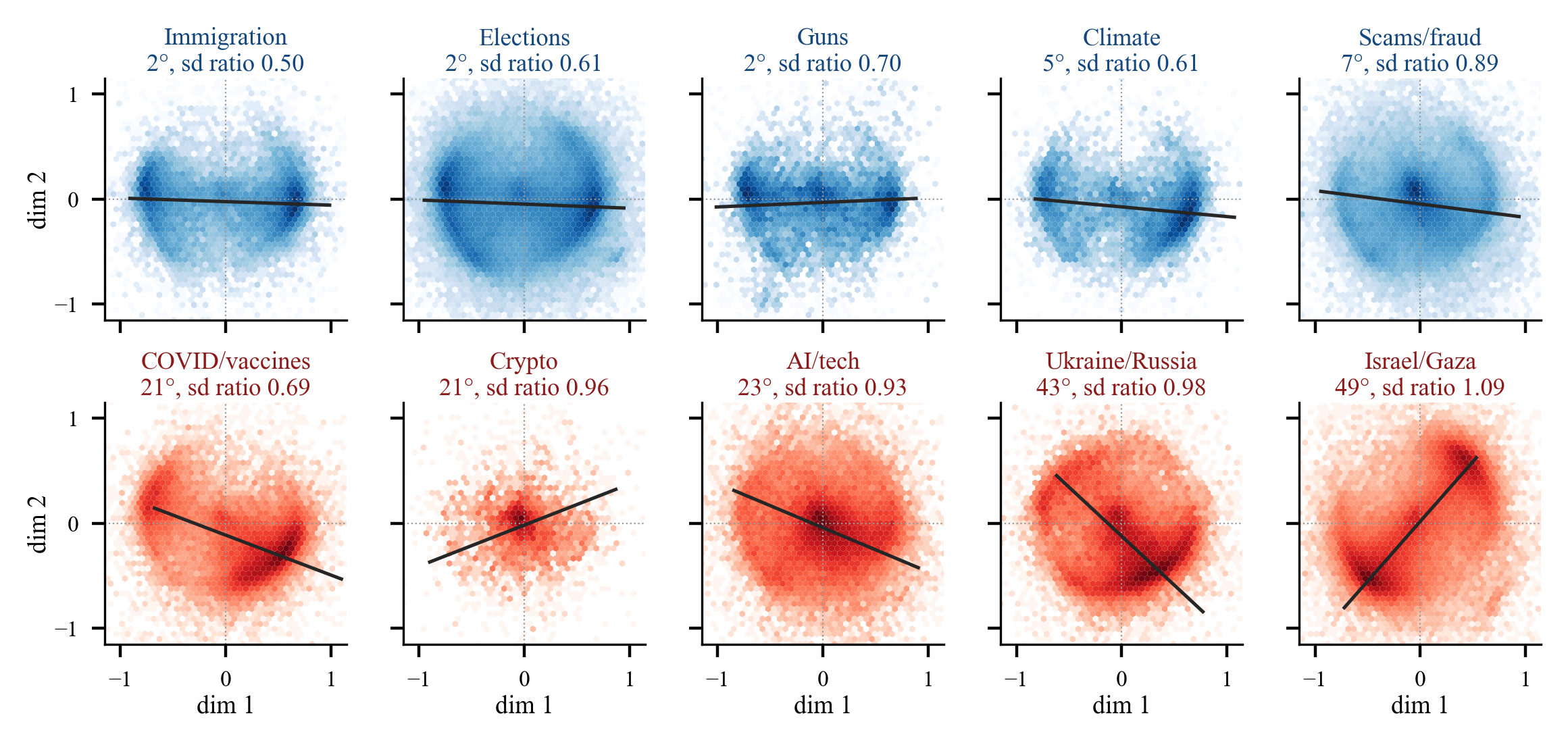}
\caption{Topic shapes in the 2-D space, on identical axes, ordered by the angle
of each topic's own principal axis (black line). If the space were
one-dimensional with noise, every topic would lie flat along dim 1. Immigration,
elections and guns do, at $2^\circ$, splitting left against right with no dim-2
structure. Ukraine/Russia and Israel/Gaza instead run diagonally at $43^\circ$
and $49^\circ$, spreading as much on dim 2 as on dim 1, their two lobes
displaced on both axes at once. ``sd ratio'' is sd$(f_2)/$sd$(f_1)$, which spans
$0.50$ to $1.09$ across the ten topics. Colour marks whether the principal axis
falls within $15^\circ$ of dim 1. Angles are unsigned, so two topics can share one and still slope opposite ways, as crypto and COVID/vaccines do at $21^\circ$.}
\label{fig:shape-all}
\end{figure*}

Keyword matching assigns 490,649 notes (21.0\%) to the 10 topics; the remaining 79.0\% are labelled ``Other'' and are excluded from the per-topic tables. The ``Other'' pool publishes at 10.42\%, close to the global rate of 10.85\%, so the assigned subset is not obviously unrepresentative on that margin. The English-centric keywords mean the assigned subset over-represents English-language political content (Section~5.4). To summarise how efficiently a group of notes converts low polarity into publication, we report a convenience index:

\begin{equation}
\text{bridgeability} = \frac{\text{publication\_rate}}{\text{mean } |f_1| + 0.01}
\end{equation}

Read it as publication earned per unit of polarisation, where a group scores high when its notes are published at a good rate while staying near the neutral point of the axis. The small constant keeps the ratio finite near the origin; the index is descriptive, not a fitted quantity, and can be sensitive to that constant for very low-polarity groups, so where a conclusion turns on it (the Greek slice in Section~4.4) we also report the raw publication rate and mean polarity separately. Table~\ref{tab:topic-bridge} gives all 10 topics, sorted by index, with their positions on both axes.

\begin{table*}[t]
\centering
\small
\begin{tabular}{lrrrrrr}
\toprule
Topic & Notes & Mean $f_1$ & Mean $|f_1|$ & Mean $f_2$ & Pub.\ rate \% [95\% CI] & Bridge. \\
\midrule
Scams/fraud & 80,515 & $-0.003$ & 0.267 & $-0.046$ & 27.2 [26.9, 27.5] & 0.98 \\
Crypto & 4,183 & $-0.015$ & 0.253 & $-0.024$ & 22.2 [21.0, 23.5] & 0.85 \\
AI/tech & 42,665 & $+0.030$ & 0.284 & $-0.056$ & 21.7 [21.3, 22.1] & 0.74 \\
Ukraine/Russia & 38,589 & $+0.077$ & 0.307 & $\mathbf{-0.195}$ & 12.6 [12.3, 13.0] & 0.40 \\
Guns & 11,640 & $-0.065$ & 0.431 & $-0.034$ & 11.3 [10.8, 11.9] & 0.26 \\
COVID/vaccines & 45,692 & $+0.212$ & 0.438 & $\mathbf{-0.194}$ & 9.0 [8.7, 9.3] & 0.20 \\
Climate & 12,247 & $+0.127$ & 0.483 & $-0.086$ & 8.7 [8.2, 9.2] & 0.18 \\
Israel/Gaza & 100,619 & $-0.094$ & 0.403 & $-0.091$ & 6.3 [6.1, 6.4] & 0.15 \\
Elections & 127,544 & $-0.002$ & 0.495 & $-0.047$ & 7.4 [7.3, 7.6] & 0.15 \\
Immigration & 26,955 & $+0.039$ & 0.530 & $-0.025$ & 6.7 [6.4, 7.0] & 0.12 \\
\bottomrule
\end{tabular}
\caption{Topic positions and bridgeability, sorted by index. ``Bridge.'' $=$ publication rate / (mean $|f_1|$ + 0.01); more negative mean $f_2$ = further out on the second axis.}
\label{tab:topic-bridge}
\end{table*}

\textbf{Reading the axes.} Dim 1 is, by construction, the dominant axis of rating disagreement, because the SVD rotation aligns it with the direction of maximum variance in the note factors. We read it as the left/right political axis because the topics that separate strongly on it are the standard partisan ones (immigration, elections, climate, COVID/vaccines), while low-conflict topics (scams, crypto) sit at its origin. This is an interpretation of the content ordering, not a measured political attribute, but external work corroborates it: \citet{bouchaud2026} find that the platform's latent factor aligns with an independently measured Left-Right scale. Guns and Israel/Gaza sit on the opposite side of the dim-1 origin  from COVID, Ukraine, climate, and immigration (Table~\ref{tab:topic-bridge}; Figure~\ref{fig:shape-all}): the opposite rater cluster finds their notes helpful, so the dominant axis does not collapse onto a single familiar left/right lineup.

On dim 2, Ukraine/Russia and COVID/vaccines occupy the most extreme positions while crypto and immigration sit nearest the origin. One reading of this ordering is that dim 2 separates topics where the divide is not only partisan (left vs right) but also epistemological (mainstream institutional or scientific consensus vs contrarian-sceptic); on that basis, we call dim 2 an institutional-trust axis. Two findings support the label. The author-factor correlation ($r = 0.358$, Section~4.3) shows that raters who contest institutional positions tend to write notes that do the same, and \citet{bouchaud2026} find that the single platform factor aligns in part with an Anti-Elite dimension (criticism of elites and institutions), which is what one would expect if a second, institutional-trust axis is being compressed into the one factor the platform fits. COVID/vaccines lies furthest from the origin in the two-axis plane (+0.212 on dim 1, $-0.194$ on dim 2), while immigration carries the highest mean absolute dim-1 polarity of any topic ($0.530$) yet sits near the dim-2 origin. So a note about COVID vaccines that achieves cross-camp publication must bridge \textit{two} independent axes of disagreement simultaneously. A rank-1 model cannot represent those axes separately, and Section~4.5 shows which notes it then cannot distinguish.

\subsection{Bridgeability and the author bias}

Topics differ substantially in how efficiently they bridge (Table~\ref{tab:topic-bridge}). Fraud-related notes  (scams, crypto fraud) bridge easily, because raters on both sides recognise fraudulent claims as harmful. Politically contested narratives (elections, Gaza) bridge poorly, because what counts as misleading is contested across camps.

We join notes to authors by matching \texttt{noteAuthorParticipantId} against the rater table, requiring that the author has given at least 5 ratings and the note has received at least 5 so both positions are estimated from real data, obtaining 1,971,666 note--author pairs from 298,833 unique authors who are also raters. Authors write notes that lean their own way on both axes: $r = 0.538$ on dim 1, $r = 0.358$ on dim 2, and 71.2\% of notes fall in the author's own camp (chance level: 50\%).

\textbf{Bridge authors.} We define bridge authors as raters strongly positioned on one side ($|f_1| > 0.3$) whose notes nonetheless stay near the centre (mean $|\text{note } f_1| < 0.1$ across at least 3 notes). Among all polarised raters ($|f_1| > 0.3$) they number 17,926 (9.4\%).

\textbf{Bridge note vocabulary.} Log-odds word frequencies between published bridge notes ($|f_1| < 0.15$) and published partisan notes ($|f_1| > 0.40$) put \textit{scam}, \textit{ethereum}, \textit{giveaway} and \textit{airdrop} at the bridge end (up to $+6$) and Serbian and French political terms and Israeli news sources, such as \textit{vucic}, \textit{climat} and \textit{ynet}, at the partisan end (down to $-4$). The bridge end is dominated by fraud vocabulary, which is consistent with bridgeability being mainly a topic effect. We did not test wording or tone. 

\subsection{Language equity: the Greek case and beyond}

Greek-script notes are detected as described in Section~3.3. Table~\ref{tab:greek-slice} and Figure~\ref{fig:greek} (Appendix~D) give results on the full dataset.

\begin{table}[t]
\centering
\small
\begin{tabular}{lrrr}
\toprule
Metric & Global & Greek & $\Delta$ \\
\midrule
Scored notes & 2,334,630 & \textbf{1,921} (0.08\%) & n/a \\
Median ratings/note & 37 & \textbf{19} & $-49\%$ \\
Pub.\ rate, crude & \textbf{10.85\%} & \textbf{7.76\%} & $-3.1$ \%-pts \\
Pub.\ rate, std. & \textbf{10.85\%} & \textbf{11.68\%} & $+0.8$ \%-pts \\
Mean $|f_1|$ & 0.3876 & \textbf{0.2785} & $-28\%$ \\
Bridgeability & 0.273 & \textbf{0.269} & $\approx$ equal \\
\bottomrule
\end{tabular}
\caption{Greek-script vs.\ global notes on the full dataset. The standardised rate reweights Greek's stratum-specific publication rates to the global distribution of ratings per note. Wilson 95\% intervals on the crude rates: global [10.81, 10.89], Greek [6.64, 9.04]. Bootstrap 95\% interval on the standardised Greek rate, 2,000 resamples: [9.96, 13.50].}
\label{tab:greek-slice}
\end{table}

\textbf{The deficit is in ratings, not in publication.} Greek notes are substantially less polarised on dim 1 (mean $|f_1|$ = 0.279 vs 0.388 globally, $-28\%$), and their crude publication rate is lower (7.76\% vs 10.85\%). That crude gap is a volume effect. Greek notes accumulate far fewer ratings, with a median of 19 versus 37 globally, and 51.0\% carry fewer than 20 ratings versus 33.2\% globally. Because publication depends steeply on rating count (Section 4.5), we standardise directly, reweighting Greek's stratum-specific publication rates to the global distribution of ratings per note. The standardised Greek rate is 11.68\%, and its interval spans the global 10.85\% (Table~\ref{tab:greek-slice}), so the crude deficit closes rather than reverses. Within strata the picture is mixed: Greek notes publish below the global rate in every stratum under 100 ratings and above it from 100 up (Figure 3).  Greek notes  reaching 100 to 499 ratings publish at 30.71\% against 17.71\% globally in the same stratum (n = 254, z = 4.5). Most of the crude Greek deficit is therefore a supply gap.  The slice is tiny for the same reason. Only 1,921 of 2,334,630 scored notes (0.08\%) are Greek-script.  For Greek-script content, Community Notes is nearly absent, and the binding constraint is the size of the rater pool, not the geometry of the bridge. The geometry is nonetheless distinctive. Greek notes average $|f_2|$ = 0.458 against 0.307 globally, a gap that holds within every rating stratum, placing them in the configuration Section~4.5 shows the rule penalises, bridging dim 1 while splitting dim 2. The standardised rate above shows no net publication cost once volume is held constant. The platform's per-group models do not reach these notes either. None of the 149 published Greek-script notes was decided by a Group model, and 1,885 of the 1,921 carry no modelling group in X's status history.

\textbf{Ten language communities.} To test whether the Greek finding generalises, we extend detection to ten language communities, combining Unicode-block membership for distinctive scripts with language-specific diacritics for Latin-script languages (full table in Appendix~D, detector details and known biases in Appendix~B; the per-language counts are approximate script-and-diacritic groupings, not validated language labels). On crude rates the pattern is not simply ``non-English $=$ disadvantaged'': high-volume communities (Portuguese 411K notes, French 295K, Turkish 134K) publish above the global average (13.2 to 14.0\% vs 10.85\%), and only two communities publish below it, Hindi (2,540 notes, 8.39\%) and Greek (1,921, 7.76\%). Those two are among the smallest but not the smallest: Korean, at 916 notes, is smaller than both and publishes at 14.19\%. Standardising every community to the global rating-count distribution resolves the apparent inconsistency. The communities whose notes attract the fewest ratings gain the most: Korean rises from 14.19\% to 22.83\% (median 15 ratings per note), Arabic from 12.24\% to 18.41\% (median 18), and Greek from 7.76\% to 11.68\% (median 19), while the two highest-volume non-Latin-script communities fall slightly (Japanese 14.70\% to 13.45\%, Chinese/CJK 14.66\% to 13.43\%). After standardisation, only Hindi remains below the global rate, at 9.29\%. Most of the difference between communities lies in how many ratings their notes attract; Hindi is the exception. With ten communities and a single snapshot, we cannot separate community size from topic mix or rater composition. Because rating count is partly downstream of publication, we report crude and standardised rates side by side. 

\subsection{What a one-axis fit cannot see}

To measure what fitting a single factor leaves out, we refit the same 212,900,053 ratings at rank 1, with identical hyperparameters, filter and held-out mask, so both fits describe the same 2,334,630 notes and score the same held-out rows (Appendix~E).

Take the quintile of notes least polarised on dim~1 and sort them into $|f_2|$ quintiles. Publication falls monotonically across them. But rating volume is a confound: notes high on dim~2 attract fewer ratings (mean 58, against 118 in the lowest quintile), and publication requires sustained activity. Imposing a minimum rating count cuts the ratio between the extreme quintiles from 17-fold to about 6-fold, and the gap persists at every threshold: 45.8\% down to 2.7\% with no threshold, 63.1\% to 6.1\% at 20 ratings, 71.9\% to 9.5\% at 50, and 71.5\% to 11.7\% at 100, quintiles recomputed inside each subset. Past the 50-rating threshold, notes that bridge dim~1 while splitting dim~2 publish at 9.5\% against 18.3\% for all notes, about half. Notes splitting dim 1 while bridging dim 2 almost never publish (0.45\%). The deployed gate on $|f_n|$ is unlikely to explain their loss, since their rank-1 factor is below average (0.24 against 0.41). Their intercept is low at both ranks (Appendix~E), so the low publication rate is not an artefact of fitting one axis; under the current intercept-based rule a two-axis fit would not publish them either. What rank 1 loses is the second coordinate, which records that raters at one end of the second axis rate these notes helpful. Whether a rule should use that information is the design choice Section~5.1 discusses.

These factors are our own estimates, since X does not release note factors; only the publication status is X's own decision.

\section{Discussion}

\subsection{Implications for system design}

Because the polarity space has at least two axes, a production model that fits one cannot represent them separately. Section~4.5 shows  that the notes that go unpublished in this configuration are  not the obviously partisan ones, which the rule already suppresses, but notes that raters across the political divide rate alike and that split instead on trust in institutions. Whether to publish such notes is the design choice this section leaves open. The direction is not obvious, since requiring cross-camp agreement on both axes is stricter than the present rule, while requiring it on either axis is more permissive. Recent proposals improve the model in other ways. \citet{goyal2026} add a per-rater quality parameter for sample efficiency and manipulation resistance, but keep the factor one-dimensional. The dimensionality of the disagreement space is a separate design choice, and the evidence here argues for revisiting it. A small core of heavy raters may set the shape of the polarity space, and if that core does not represent the wider community, the factor estimates and the publication decisions they drive may reflect the preferences of an unrepresentative minority. 

\subsection{The author bias problem}

Because authors write to their own side ($r = 0.538$ on dim 1, $r = 0.358$ on dim 2), the note corpus leans toward whichever side produced it, on both the political and the institutional-trust axis. Cross-camp notes take either a bridge author (about 9.4\% of polarised raters) or an author near the centre. Recruiting for raw participation instead of a spread of viewpoints could deepen the tilt, and rater concentration compounds it. As the busiest raters accumulate the firmest factor estimates, their bias may weigh most on which notes clear the bridge.

\subsection{Language equity implications}

Uptake of Community Notes varies widely across languages \citep{stewart2025}, and small language communities are the ones it reaches least (Section~4.4). Greek-script content is the clearest case in our data. The Greek-language public sphere, Cyprus included, has documented exposure to online disinformation \citep{giomelakis2024}. The inequity is emergent rather than designed, and Section 4.4 finds its cause upstream of the bridging rule. A small language supplies too few raters for notes to reach the volume the publication threshold requires, while the notes that do reach it publish at or above the global rate. Recruiting raters within specific languages addresses the actual bottleneck, and machine translation could widen the effective rater pool for a given note. Lowering the publication bar for small rater pools, which the unadjusted publication rates on their own would suggest, is not supported by the standardised comparison, and it would be the riskiest option in precisely the small, cohesive communities where a coordinated group is cheapest to assemble \citep{truong2025}.

\subsection{Limitations}

\begin{enumerate}
\item \textbf{Dimensionality and reproducibility.} X retains only its most recent public data release, which has since shrunk (the 2026-07-05 snapshot holds 55.7M usable ratings against the 212.9M behind our descriptive fit), so the full-scale fit cannot be rerun from the current public release, though the rank comparison up to $K = 2$ was run on the 2026-06-14 ratings themselves (Section~4.1). The qualitative conclusion is robust, as three separate fits, on 212.9M ratings (descriptive, 2026-06-14), 55.7M (held-out, 2026-07-05), and 45.9M (transfer, a different topic subset refit locally on CPU), all recover a second predictive axis.

\item \textbf{Pipeline scope.} Publication status reflects X's full pipeline (Section~2.1), while our factors and intercepts come from the base model on a different scale, so we compare notes with each other rather than against X's thresholds.

\item \textbf{Language detection.} Our detection method identifies script and diacritics, not language. Greek-script detection captures Greek and Cypriot content but not Greek-topic content written in English. Among Latin-script languages, the diacritic rules overlap, so Portuguese and French mutually misclassify, and Spanish, which has no detector of its own, is largely absorbed into the Portuguese bucket. The per-language counts (Section~4.4, Appendix~B) are approximate.

\item \textbf{Topic assignment.} Our keyword matching is English-centric. Non-English notes on the same political topics are classified as ``Other,'' causing the topic analysis to underestimate political polarisation in non-English communities.

\item \textbf{Temporal coverage.} The analysis is a cross-sectional snapshot (June 2026). The evolution of the second axis and of language coverage gaps over time is not characterised.

\item \textbf{Axis interpretation.} We infer both semantic labels from topic structure. We have no direct measures of rater ideology or trust in institutions, and the sign of each axis is arbitrary. External work partially supports the dim-1 reading (\citet{bouchaud2026} find the platform factor aligns with a Left-Right scale), and the cross-topic transfer test (Section~4.1) shows a second rater dimension that generalises across topics, but that test does not identify the transferring axis with dim 2, and the institutional-trust label remains our own interpretation. Two rater-level alternatives that would also transfer across topics, rater leniency and a curved dependence on dim 1, are not excluded by the transfer test. Among raters with 50 or more ratings, dim 2 correlates $-0.23$ with the rater intercept and $+0.26$ with the square of dim 1, and the two together explain 9.6\% of its variance; among notes with 50 or more ratings the corresponding figures are $-0.12$, $+0.10$ and 1.8\%. Rater positions on dim 2 also vary with X's modelling groups, assigned by region, country or language, which explain 7.1\% of dim-2 variance among raters with 50 or more ratings against 3.9\% for dim 1, so part of the second axis may be regional or linguistic. Similarly, the bimodality coefficients are activity-dependent (Section~4.2). Inactive raters are shrunk towards the origin, so full-pool values understate camp structure, and we report $BC$ as a function of rating count.

\item \textbf{Author-exposure confound.} The author-note factor correlation (Section~4.3) is measured on users who both write and rate. Because a rater's factor is estimated partly from the notes they chose to rate, and authors tend to write and rate in the same topic areas, part of the correlation may reflect shared topical exposure rather than a stable cross-topic viewpoint. We report it as an association, not as a causal author effect.
\end{enumerate}

\section{Conclusion}

Community Notes runs on a one-dimensional model of disagreement, and at full scale that model is incomplete. Alongside the left/right axis sits a second, largely independent axis that we read, provisionally, as trust in institutions. A rule that cannot see the second axis cannot tell notes supported by raters at one end of the second axis from notes of generally lower appeal, and those are the notes that cross the political divide.  The smallest language communities lose notes for a different reason: nothing in the rule discriminates against them, but the raters who could close the bridge are not there to rate.

Two changes follow. Give the bridging model more than one dimension, and recruit raters in the languages the current system reaches least. Where the pool is too small for the bridge to close, a third is needed, fact-checking that does not depend on bridging. Bridging-based ranking is now proposed well beyond fact-checking \citep{ovadya2023}, and any system that models disagreement with fewer axes than the data hold risks inheriting this blind spot. 

\section*{Data, Ethics, and Reproducibility}

\textbf{Data.} This study uses only the public X Community Notes data release (note text, ratings, and note-status history), retrieved from X's public data endpoint. Contributor identifiers are pseudonymous participant IDs assigned by the platform; we did not attempt to deanonymise them or link them to external data. We analyse records in aggregate and report no individual-level content beyond text already published on the platform.

\textbf{Ethics.} The work is a secondary analysis of publicly released, pseudonymised platform data with no interaction or intervention with human participants, and so does not constitute human-subjects research under common institutional-review criteria; no ethics-board approval was required. Use of the public data release follows the platform's terms.

\textbf{Reproducibility.} The model is the platform's published bridging matrix factorisation, refit at full scale; all hyperparameters appear in Section~3. Analysis code and the result artefacts from which every reported number was read are available in an OSF repository: \url{https://osf.io/n5w4z/?view_only=d1de855b3fb04977a7f35277d6a59e68}. One limitation is inherent to the source: X's public release retains only the most recent snapshot, so the exact 2026-06-14 rating file is no longer retrievable (Sections~4.1 and~5.4).

\textbf{Use of generative AI.} Generative AI assistance (Anthropic Claude) was used in preparing this manuscript: drafting and revising prose, generating and reviewing analysis and figure code, and running consistency and source-verification checks against the computational artefacts. The research questions, study design, choice of analyses and interpretation of results are the authors' own; every reported number was traced to the artefact that produced it, and the authors take full responsibility for the content. Generative AI is not listed as an author.

\section*{Acknowledgments}
This work has been funded by the European Union under the project: MedDMO II (Grant Agreement no.~101226175).

{\small
\bibliography{references}
}

\clearpage
\appendix

\renewcommand{\topfraction}{0.9}
\renewcommand{\dbltopfraction}{0.9}
\renewcommand{\bottomfraction}{0.5}
\renewcommand{\textfraction}{0.15}
\renewcommand{\floatpagefraction}{0.5}
\renewcommand{\dblfloatpagefraction}{0.5}
\setcounter{topnumber}{3}
\setcounter{dbltopnumber}{3}
\setcounter{totalnumber}{4}
\raggedbottom
\makeatletter
\setlength{\@fptop}{0pt}
\setlength{\@fpsep}{12pt}
\setlength{\@fpbot}{0pt plus 1fil}
\setlength{\@dblfptop}{0pt}
\setlength{\@dblfpsep}{12pt}
\setlength{\@dblfpbot}{0pt plus 1fil}
\makeatother

\section{Validation of the Refit}

Logistic regression on note status, using each note's fitted intercept $i_n$ and factor $f_1$ from the earlier one-dimensional fit (Section~3.2) as predictors:

\begin{itemize}
\item AUC $= \textbf{0.993}$ for CURRENTLY\_RATED\_HELPFUL vs NOT\_HELPFUL
\item AUC $= \textbf{0.880}$ for CURRENTLY\_RATED\_HELPFUL vs all other statuses (which include the large NEEDS\_MORE\_RATINGS pool)
\item The separation is carried almost entirely by the intercept $i_n$ (AUC $0.993$ on its own); $f_1$ alone does not separate the two classes (AUC $0.44$, below the 0.5 chance level), consistent with the factor's role in the publication rule being to gate agreement across camps rather than to predict helpfulness.
\item We take the first contrast as primary because it is the cleaner two-class comparison, and report the all-statuses figure beside it so the number is not over-read.
\end{itemize}

Our re-estimated note intercepts closely match the official model's published decisions. This supports the refit, but does not mean we recover X's model exactly. 

\section{Topic and Language Detectors}

Table~\ref{tab:keyword-patterns} lists the ten patterns exactly as applied, with the two matching modes and the ordering that decides which bucket a note falls into.

\begin{table*}[tp]
\centering
\small
\begin{tabular}{lp{0.62\textwidth}l}
\toprule
Topic & Prefix, case-insensitive & Whole word, case-sensitive \\
\midrule
Elections & elect(?!r)$|$ballot$|$voter$|$voting$|$democrat$|$republican$|$congress$|$president & \\
COVID/vaccines & covid$|$coronavirus$|$vaccine$|$vaxx?$|$pfizer$|$moderna$|$pandemic & mRNA \\
Climate & climat$|$global warm$|$carbon$|$emission$|$fossil fuel$|$renewable$|$greenhouse & \\
Ukraine/Russia & ukraine$|$russia$|$zelensky$|$putin$|$kyiv$|$donbas$|$crimea & NATO \\
Israel/Gaza & israel$|$gaza$|$hamas$|$palestin$|$netanyahu$|$west bank$|$ceasefire & IDF \\
Immigration & immigr$|$migrant$|$border$|$asylum$|$deporta$|$undocumented & CBP, ICE \\
Guns & gun$|$firearm$|$2nd amendment$|$AR-15$|$mass shoot$|$gun control & NRA \\
Scams/fraud & scam$|$fraud$|$phish$|$ponzi$|$fake$|$hoax$|$impersonat & \\
Crypto & crypto$|$bitcoin$|$ethereum$|$blockchain$|$binance$|$coinbase & NFT, FTX \\
AI/tech & artificial intelligence$|$chatgpt$|$openai$|$deepfake & AI \\
\bottomrule
\end{tabular}
\caption{Topic keyword patterns, as applied. Stems match as prefixes with a leading word boundary, case-insensitively; short acronyms match as whole words, case-sensitively, since as case-insensitive prefixes ICE would match ``iceberg'' and AI would match ``aid''. The lookahead in elect(?!r) admits elect, elected, election, electoral and electorate while excluding the electr- family. Matching runs on the raw summary, URLs included, so a cited domain can drive the assignment. First match wins, so the buckets are ordered rather than independent: a Ukraine note that also mentions a president is assigned to Elections. Unmatched notes are labelled ``Other''.}
\label{tab:keyword-patterns}
\end{table*}

\textbf{Language detection.} Detection combines two heuristics: Unicode-block membership for languages with distinctive scripts (Arabic, Hindi/Devanagari, Chinese/CJK, Japanese, Korean, and Greek), and the presence of language-specific diacritics for Latin-script languages (Turkish \u{g}/\i/\c{s}, Portuguese \~{a}/\~{o}/\c{c} and accented vowels, French \c{c}/\oe\ and accented vowels, German \"{a}/\"{o}/\"{u}/\ss). A note is assigned to a language if its summary contains any matching character, so the buckets are not mutually exclusive. The detector is deliberately simple and carries two known biases: the Latin-script rules key on diacritics that several languages share, so Portuguese and French overlap and mutually misclassify; and Spanish, which has no detector of its own and shares its accented vowels (\'{a}, \'{e}, \'{i}, \'{o}, \'{u}) with the Portuguese rule, is largely absorbed into the Portuguese bucket and inflates it. The per-language counts are approximate script-and-diacritic groupings, not validated language labels, but the individual figures carry measurement error. 
\section{Robustness of the Two-Dimensional Reading}

\textbf{Bimodality by activity filter.} Table~\ref{tab:bc-activity} gives Sarle's coefficient by activity level.

\begin{table*}[tp]
\centering
\small
\begin{tabular}{lrlrrlr}
\toprule
Activity filter & Raters $n$ & BC d1 [95\% CI] & BC d2 & Notes $n$ & BC d1 [95\% CI] & BC d2 \\
\midrule
all & 1.07M & 0.378 [0.372, 0.384] & 0.198 & 2.33M & 0.395 [0.393, 0.397] & 0.221 \\
$\geq$20 ratings & 0.88M & 0.488 [0.478, 0.497] & 0.260 & 1.56M & 0.504 [0.501, 0.507] & 0.322 \\
$\geq$50 ratings & 0.60M & \textbf{0.597} [0.588, 0.604] & 0.360 & 0.97M & 0.555 [0.553, 0.556] & 0.338 \\
$\geq$100 ratings & 0.38M & \textbf{0.641} [0.628, 0.651] & 0.384 & 0.56M & \textbf{0.597} [0.596, 0.599] & 0.340 \\
\bottomrule
\end{tabular}
\caption{Sarle's bimodality coefficient by activity filter, for rater and note factors, with bootstrap 95\% intervals over 300 resamples. The rater dim-1 coefficient clears the 0.555 bimodality threshold at 50 or more ratings; the note coefficient only at 100 or more, its interval at 50 straddling the threshold. Dim 2 lies below it at every filter. For notes, the filter counts ratings per note in the fitted set, as in Section~4.5.}
\label{tab:bc-activity}
\end{table*}

\section{Participation and Language Coverage}

\textbf{Rater participation inequality.} The rating distribution across raters is heavily concentrated: the Gini coefficient over ratings per rater is \textbf{0.718} (0 would mean every rater contributes equally, 1 that a single rater does all the rating), with the top 1\% of raters contributing \textbf{23.1\%} of all ratings and the top 10\% contributing \textbf{61.5\%}. The estimates for the most active raters rest on thousands of ratings, while most raters contribute few. This has two implications: (i) the polarity space is disproportionately shaped by a small core of highly active raters, whose viewpoints may not represent the broader community; and (ii) publication decisions for notes that attract ratings from low-activity raters are more susceptible to random noise in the factor estimates.

\textbf{Multi-language coverage} (Table~\ref{tab:multilang}). Because publication depends steeply on how many ratings a note accumulates (Section 4.5) and the communities differ several-fold in ratings per note, Table~\ref{tab:multilang} reports each community's publication rate both crude and standardised to the global distribution of ratings per note.

\begin{table*}[tp]
\centering
\small
\begin{tabular}{lrrrlrlr}
\toprule
Language & $N$ notes & Med.\ rat. & Pub.\ rate & 95\% CI & Pub.\ std. & 95\% CI & Bridge. \\
\midrule
\textbf{GLOBAL} & \textbf{2,334,630} & \textbf{37} & \textbf{10.85\%} & [10.81, 10.89] & \textbf{10.85\%} & (reference) & \textbf{0.273} \\
Portuguese & 411,056 & 33 & 13.26\% & [13.16, 13.36] & 13.74\% & [13.63, 13.84] & 0.305 \\
French & 295,419 & 34 & 13.96\% & [13.84, 14.09] & 14.35\% & [14.22, 14.48] & 0.320 \\
Japanese & 180,687 & 43 & 14.70\% & [14.54, 14.86] & 13.45\% & [13.30, 13.59] & 0.459 \\
Chinese/CJK & 179,783 & 42 & 14.66\% & [14.49, 14.82] & 13.43\% & [13.29, 13.57] & 0.457 \\
Turkish & 134,332 & 35 & 13.21\% & [13.03, 13.39] & 13.34\% & [13.17, 13.52] & 0.282 \\
German & 62,338 & 30 & 12.13\% & [11.88, 12.39] & 13.20\% & [12.92, 13.47] & 0.251 \\
Arabic & 12,652 & 18 & 12.24\% & [11.68, 12.82] & 18.41\% & [17.60, 19.23] & 0.299 \\
Hindi/Devanagari & 2,540 & 31 & 8.39\% & [7.37, 9.53] & 9.29\% & [8.17, 10.41] & 0.187 \\
Greek & 1,921 & 19 & 7.76\% & [6.64, 9.04] & 11.68\% & [9.96, 13.50] & 0.269 \\
Korean & 916 & 15 & 14.19\% & [12.08, 16.60] & 22.83\% & [19.65, 25.98] & 0.428 \\
\bottomrule
\end{tabular}
\caption{Multi-language coverage by detected language community. ``Pub.\ std.'' is the crude rate standardised to the global distribution of ratings per note (six strata: 5-9, 10-19, 20-49, 50-99, 100-499, 500+). Intervals on the crude rates are Wilson; those on the standardised rates are bootstrap percentile over 2,000 resamples of notes within each community. Greek is the only community whose standardised interval spans the global rate, and Hindi the only one resolved below it.}
\label{tab:multilang}
\end{table*}

\begin{figure*}[tp]
\centering
\includegraphics[width=0.6\textwidth]{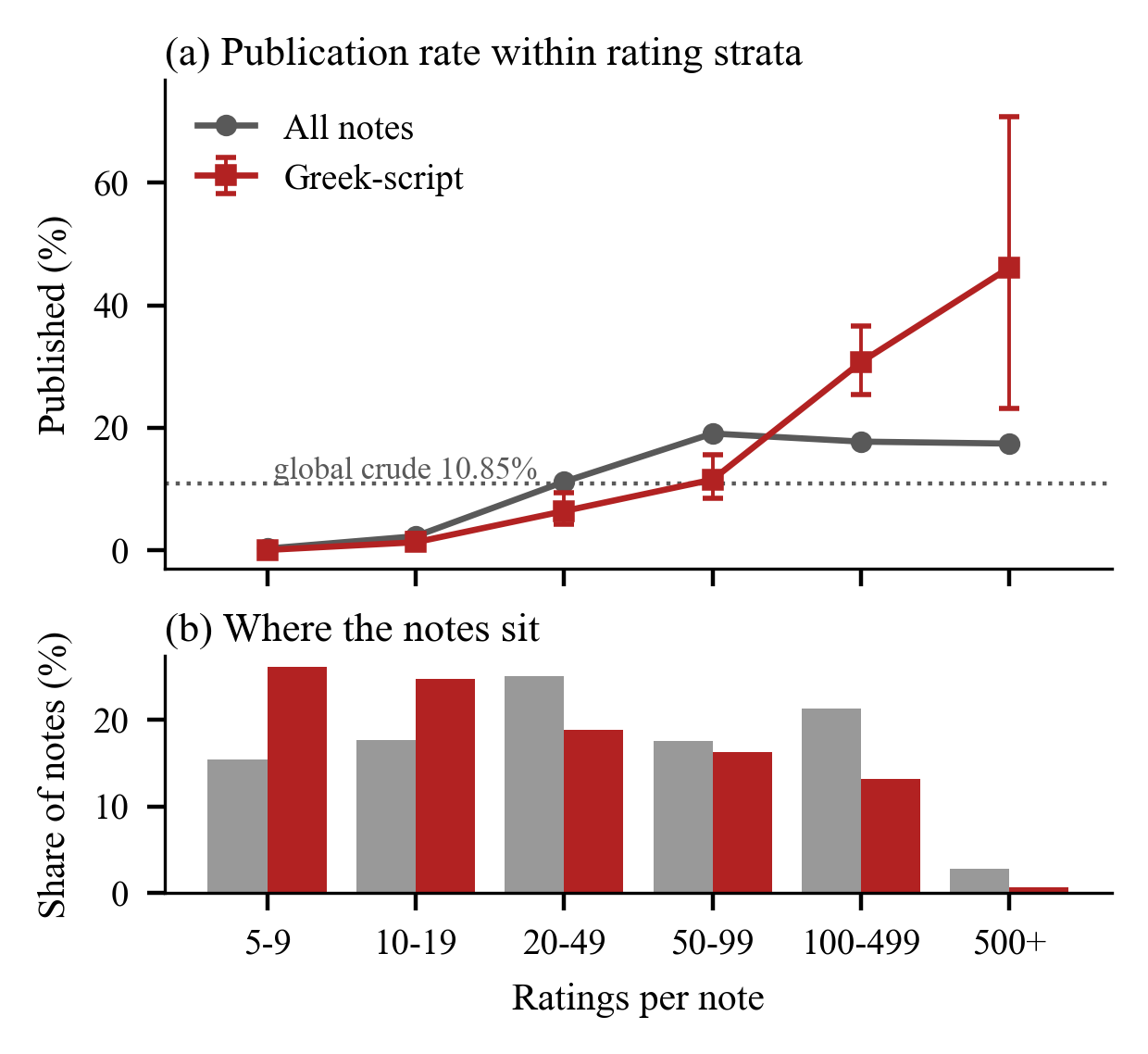}
\caption{The Greek deficit is in rating supply, not publication. (a) Publication rate by ratings-per-note stratum, Greek-script (n = 1,921) against all notes, with Wilson 95\% intervals on the Greek points: Greek sits below the global rate in every stratum under 100 ratings and above it beyond, though the top stratum holds only 13 notes. (b) Greek mass sits in the low-rating strata, where few notes of any kind publish, which is what pulls the crude Greek rate to 7.76\%. Standardised to the global rating-count distribution it is 11.68\%, against a global 10.85\% (Table~\ref{tab:greek-slice}).}
\label{fig:greek}
\end{figure*}

\section{What a rank-1 fit does with the second axis}

We refit the 2026-06-14 ratings at $K = 1$ under the settings used for the rank-2 fit (lr 0.05, $\lambda_i$ = 0.15, $\lambda_f$ = 0.03, 300 epochs, seed 0, 2\% held out) on the identical filtered set, 212,900,053 training ratings over 2,334,630 notes and 1,067,909 raters, so the two fits score the same held-out rows. Held-out RMSE is 0.3022 after 300 epochs; the early-stopped comparison in Section~4.1 reached 0.2936 for $K = 1$ on the same snapshot.

Writing $g$ for the note factor of the rank-1 fit, and $|f_1|$ and $|f_2|$ for the rank-2 factors, a regression of the rank-1 factor magnitude on both rank-2 axis magnitudes gives $|g| = 0.046 + 0.791|f_1| + 0.185|f_2|$ ($R^2 = 0.30$), with standardised coefficients 0.533 on $|f_1|$ and 0.124 on $|f_2|$. The second-axis coefficient is small but clearly non-zero, so a single coordinate partly tracks the second axis. Taken alone, $|f_2|$ accounts for 2.1\% of the variance in the rank-1 factor magnitude.

In the quintile least polarised on dim 1 the rank-1 note intercept runs 0.793, 0.759, 0.701, 0.647, 0.581. The rank-2 intercept falls the same way, 0.801, 0.766, 0.693, 0.627, 0.563, so the drop does not come from fitting one axis. Among notes with 50 or more ratings in that quintile, those in the top $|f_2|$ quintile carry a mean rank-1 factor magnitude of 0.24, against 0.41 for all notes past 50 ratings, and 10.6\% of them fall in the top fifth of $|g|$. The deployed gate on $|f_n|$ therefore does little to these notes, and a one-axis rule sees a weakly polarised note of lower helpfulness where the rank-2 fit sees a note that raters at one end of the second axis support.

\end{document}